\documentclass[aps,pre,twocolumn,nofootinbib,10pt]{revtex4-1}

\usepackage[utf8]{inputenc}
\usepackage{mathtools}
\usepackage[T1]{fontenc}
\usepackage{mathptmx}
\usepackage[utf8]{inputenc}
\usepackage{gensymb}
\usepackage{graphicx}
\usepackage{subcaption}
\usepackage{amsmath}
\usepackage{amssymb}
\usepackage{dcolumn}
\usepackage{bm}
\usepackage{xr}
\usepackage{cleveref}
\usepackage{soul}
\usepackage{color}
\usepackage{ulem}
\usepackage[usenames,dvipsnames]{xcolor}

\usepackage{graphicx}
\usepackage{dcolumn}
\usepackage{bm}

\usepackage[utf8]{inputenc}
\usepackage{mathtools}
\usepackage[T1]{fontenc}
\usepackage{mathptmx}
\usepackage[utf8]{inputenc}
\usepackage{graphicx}
\usepackage{amsmath}
\usepackage{amssymb}
\usepackage{dcolumn}
\usepackage{bm}
\usepackage{xr}
\usepackage{cleveref}
\usepackage{soul}
\usepackage{color}
\usepackage{ulem}
\usepackage[usenames,dvipsnames]{xcolor}

\usepackage[utf8]{inputenc}
\usepackage{mathtools}
\usepackage[T1]{fontenc}
\usepackage{mathptmx}
\usepackage[utf8]{inputenc}
\usepackage{graphicx}
\usepackage{subcaption}
\usepackage{amsmath}
\usepackage{amssymb}
\usepackage{dcolumn}
\usepackage{bm}
\usepackage{xr}
\usepackage{makecell}
\usepackage{cleveref}

\begin{document}

\title{Simple Calibration of Block Copolymer Melt Models}
\author{Artem Petrov}
 \email{aipetrov@mit.edu}
	\affiliation{Department of Chemical Engineering, Massachusetts Institute of Technology, Cambridge, Massachusetts 02139, United States}
 \author{Hejin Huang}
	\affiliation{Department of Materials Science and Engineering, Massachusetts Institute of Technology, Cambridge, Massachusetts 02139, United States}
 \author{Alfredo Alexander-Katz}
\email{aalexand@mit.edu}
	\affiliation{Department of Materials Science and Engineering, Massachusetts Institute of Technology, Cambridge, Massachusetts 02139, United States}
	\date{\today}
	
\begin{abstract}

According to the universality hypothesis, the phase behavior of different block copolymer melt models having fixed composition depends solely on two parameters: the invariant chain length $\bar{N}$ and the effective interaction parameter $\chi N$. If models behave universally, they can be compared to each other and can predict experiment quantitatively. Here, we present a simple way to achieve this universality for coarse-grained models. Our method relies on the properties of the monomer interaction potential energy $z$  distribution. In particular, models having near-symmetric $z$-distributions exhibit universal phase behavior using the standard linear definition of the Flory-Huggins parameter $\chi\propto\alpha$, where $\alpha = \epsilon_{AB}-(\epsilon_{AA}+\epsilon_{BB})/2$, and $\epsilon_{xy}$ is the interaction energy between monomers of type $x$ and $y$. Previously, universality had been achieved using a nonlinear $\chi(\alpha)$ function which is difficult to obtain and interpret physically. The main parameter controlling the symmetry of the $z$-distribution is the monomer density $\rho$. Above certain $\rho$, models have symmetric $z$-distributions, and their order-disorder transition points follow the universal curve predicted by Fredrickson-Helfand theory in the experimentally relevant $\bar{N} > 10^2$ range.
On the other hand, low-$\rho$ models exhibit skewed $z$-distributions, and the simple $\chi\propto\alpha$ formula is no longer universally applicable to them. Our results can be used for correct block copolymer model building leading to a simple and direct comparison of simulations to experiments, which will facilitate the screening of new block copolymer morphologies and support materials design.

\end{abstract}

\maketitle

Melts consisting of block copolymers are able to form ordered microphases in which the monomers of different types are distributed periodically in space. Such structures have immense importance in nanotechnology applications, especially in nanophotonics, green plastics, and nanolithography \cite{sinturel2015high}. The simplest example of a block copolymer is a symmetric diblock copolymer that consists of two A and B blocks having equal volume fractions. Such polymers are able to self-assemble into nanometer-scale periodic lamellar structures at the so-called order-disorder transition (ODT) temperature. Predicting the location of the ODT point from the physico-chemical properties of polymers, which is crucial for building a phase diagram of any block copolymer material, still remains an unsolved problem despite the simplicity of this polymer system and five decades of research \cite{helfand1975block}.

The measure of dissimilarity between the monomers of type A and B is the Flory-Huggins parameter $\chi$; it characterizes the positive enthalpic contribution to the free energy of a block copolymer melt \cite{leibler1980theory}. Leibler's self-consistent (mean) field theory (SCFT) predicted that the product of $\chi$ at ODT and the chain degree of polymerization $N$ equals to a single number $(\chi N)_\text{ODT}=10.495$ \cite{leibler1980theory}. However, this theory disregarded density fluctuations at scattering vectors different from the instability vector near the ODT point; as a result, SCFT is expected to work only for infinitely long polymers \cite{grzywacz2007renormalization}.

We can characterize how close a certain system is to the mean-field SCFT regime by the invariant chain length $\bar{N}=(\rho R^3/N)^2$, which plays the role of the Ginzburg parameter \cite{medapuram2015universal}, where $\rho$ is the polymer concentration and $R$ is the mean end-to-end distance of a copolymer. After the seminal work of Leibler, a number of researchers attempted to take density fluctuations at ODT into account more correctly. Following Brazovskii \cite{brazovskiǐ1975phase}, the first theory that included fluctuations was developed by Fredrickson and Helfand (FH) \cite{fredrickson1987fluctuation} and corrected by Olvera de la Cruz (OC) and co-workers \cite{mayes1991concentration}. They found that the $(\chi N)_\text{ODT}$ value for finite symmetric diblock copolymers should be larger than the SCFT prediction:  
\begin{equation}
\label{eq:1}
    (\chi N)_\text{ODT}=10.495+41.0\bar{N}^{-1/3}
\end{equation}
The FH theory was supported by several experimental works \cite{bates1990fluctuation,bates1988fluctuation,rosedale1995order}. 

In the last few decades, with the advent of fast computers, a lot of attention in polymer science was attracted to computer simulations. Modeling provides researchers with an inexpensive way to study the complex phase behavior of block copolymers at the most detailed level. However, the biggest challenge arising when one tries to compare the results of a simulation to an experiment is the difficulty of mapping the simulation parameters onto experimentally measurable quantities. For instance, the $\chi$ parameter, which is utilized in experiments to control the miscibility of A and B blocks, is not used as an input parameter in most simulations. In field-theoretic modeling, a "bare" interaction parameter $\chi_b$ is adjusted; in particle-based simulations, the exchange energy $\alpha = \epsilon_{AB}-(\epsilon_{AA}+\epsilon_{BB})/2$ is varied. The $\chi$ parameter is a monotonically increasing function of either $\chi_b$ or $\alpha$; however, there exists a plethora of mappings between these quantities \cite{groot1997dissipative,medapuram2015universal,qin2012fluctuations,willis2019calibration,vorselaars2015field}. The most widely used relation is a simple linear approximation $\chi=z_{mod}\alpha/k_BT$, where $T$ is temperature and $z_{mod}$ is a model-dependent prefactor characterizing the monomer coordination number \cite{glaser2014universality,medapuram2015universal,morse2009chain,groot1997dissipative}. This natural approximation has a clear physical meaning: the free energy of interaction per monomer is proportional to the potential energy of a monomer-monomer contact. Another great virtue of this relation is its direct applicability to various types of block copolymers.

However, it turned out that this approximation had a serious disadvantage. Several simulation works \cite{zong2013fluctuation,glaser2014universality,medapuram2015universal} pointed out that if the $\chi\propto \alpha$ expression was assumed, the $(\chi N)_\text{ODT}$ values depended not only on $\bar{N}$ but also on other model details that do not have strict correspondence to the characteristics of real polymer systems. In other words, the $\chi\propto \alpha$ approximation yielded a so-called "nonuniversal" behavior of polymer models at ODT, which hampered the comparison of simulations to experiments. This effect was observed for the values of $\bar{N}$ that are usually achieved experimentally: $10^2<\bar{N}<10^4$.

The solution to this problem was found by Morse and co-workers \cite{glaser2014universality,medapuram2015universal} who showed that a \textit{nonlinear} $\chi(\alpha)$ relation leads to the universal $(\chi N)_\text{ODT}(\bar{N})$ dependency for a variety of particle-based models. Moreover, they demonstrated that the $(\chi N)_\text{ODT}$ values were substantially larger than the FH prediction in the range $10^2<\bar{N}<10^4$ (Eq. \ref{eq:2}). 
\begin{equation}
\label{eq:2}
    (\chi N)_\text{ODT}=10.495+41.0\bar{N}^{-1/3}+123.0\bar{N}^{-0.56}
\end{equation}
Later, Matsen and coworkers found that the same formula predicted well the $(\chi N)_\text{ODT}$ values in field-theoretic simulations if $\chi$ was allowed to be a nonlinear function of $\chi_b$ \cite{beardsley2019computationally,beardsley2019calibration}.

This nonlinear $\chi(\alpha)$ relation is model-specific and can be found from mapping the structure factor of the disordered phase of a diblock copolymer melt model to the predictions of renormalized one-loop (ROL) theory \cite{grzywacz2007renormalization,qin2009renormalized,qin2011renormalized}. One of the biggest downsides of this mapping is that it is rather difficult to perform: it requires simulation of the systems and the measurement of the structure factor at different values of $\alpha$ as well as solving complex integrals numerically in addition to the determination of the $z_{mod}$ parameter described above. Moreover, strictly speaking, ROL theory has only been developed for binary blends and diblock copolymers; it is still an open question whether the current form of the theory is applicable to other types of block copolymers which, in turn, are of much interest and technological importance. In addition, the physical nature of the nonlinear terms in $\chi(\alpha)$ is still unclear. As a result, the majority of simulation works currently operate with the simple $\chi\propto\alpha$ expression \cite{huang2019dissipative,jiao2016computer,petrov2022phase,beardsley2022well,vorselaars2015field,zong2015order,gavrilov2020polymerization} despite the issues with unphysical model-dependent behavior of simulated systems and the lack of understanding of when this approximation is applicable. Furthermore, modern experimental research groups, which do not have expertise in the peculiarities of ROL theory, tend to perform simulations of block copolymer melts to gain more detailed insight into the observed phenomena \cite{arora2016broadly}. This emphasizes the need for the method of deriving the $\chi(\alpha)$ function which (i) is as simple and general as possible and (ii) yields the $(\chi N)_\text{ODT}$ values dependent only on $\bar{N}$ and not on other parameters of the models used.

Here, we propose a solution to this problem. We prove that the simplest $\chi\propto\alpha$ approximation leads to the universal behavior of block copolymer melt models if the effective coordination number $z(N)$ has a (quasi) \textit{symmetric} distribution around its mean. $z(N)$ represents the dimensionless potential energy exerted on a monomer by other polymer chains of length $N$ \cite{morse2009chain}. 
In turn, the $z(N)$-distribution is nearly symmetric if model density is large enough. 
Moreover, we found that the universal $(\chi N)_\text{ODT}(\bar{N})$ dependency exhibited by such models has the functional form of the FH-OC model and is rather close to the original prediction by Fredrickson and Helfand (Eq. \ref{eq:1}) and not to the expression of Morse et al. (Eq. \ref{eq:2}) for $\bar{N}>10^2$. Utilizing these models in the future research on phase separation of block copolymers will not require the complex nonlinear renormalization of $\chi$ and will give physically correct universal results even for polymers with moderate $\bar{N}$. As a result, it will be possible to determine the experimental $\chi(T)$ function directly from this $(\chi N)_\text{ODT}(\bar{N})$ dependency \cite{bates1990fluctuation} and, therefore, to compare the simulations of such models to experiments easily for any type of block copolymers.

We modeled diblock copolymer melts using dissipative particle dynamics (DPD) and the bead-and-spring representation of polymers. Soft-core repulsion force acted between all beads and defined the A-B repulsion energy $\alpha$ (Eq. \ref{eq:33}). Chain connectivity was modeled by the harmonic spring force (Eq. \ref{eq:3}).
\begin{equation}
\label{eq:3}
\bold{F}_{ij}^b=-K(r_{ij}-r_0) \frac{\bold{r}_{ij}}{r_{ij}}
\end{equation}

Here, $\bold{r}_{ij}$ is the vector between bonded beads $i$ and $j$, and $K$ is the bond stiffness. Supporting Information (SI), section 1 contains the full description of the simulation procedure. We chose high $K=100$ to model freely-jointed chains (FJC)  \cite{gavrilov2023effect,huang2019dissipative} and varied $K$ from $K=0.867$ to $K=4.0$ to model "Gaussian chains": soft-spring models having $r_0=0$, which are traditionally used in particle-based simulations \cite{groot1998dynamic,medapuram2015universal}. In this paper, a "model" is a set of interaction potential parameters and the values of $N$ and $\rho$. We constructed 32 DPD models of symmetric diblock copolymers to study their phase behavior at experimentally relevant values of invariant chain length $10^2<\bar{N}<10^4$ (Table \ref{table}). We calculated $\bar{N}$ according to the definition $\bar{N}=N(\rho b^3)^2$, where $\rho$ is the number of beads per unit volume and $b$ is statistical segment length determined from extrapolating the radius of gyration $R_g^2 = N b^2 / 6$ to $N\to\infty$ (see SI, section 2). The full set of parameters for each model is given in Table \ref{table2}. To support our conclusions, we also included previously published data for non-DPD models of symmetric diblock copolymers. First, we reviewed the results of the field-theoretic simulations (FTS) \cite{beardsley2019calibration,beardsley2019computationally}, which used the standard Gaussian model of polymer chains \cite{matsen2001standard}. Second, we included the data from ref. \cite{willis2019calibration} for the lattice models with a maximum single site occupancy $Z=5$ (Table \ref{table}).

\begin{table}
 \caption{Parameter ranges of the studied models: freely-jointed chains (FJC), Gaussian chains (G), multiple-occupancy lattice models \cite{willis2019calibration}, and field-theoretic simulations (FTS) \cite{beardsley2019calibration,beardsley2019computationally}.}
 \label{table}
\begin{tabular}{@{}llllll@{}}
    \hline
    Model & $\rho$ & $N$ & $K$ & $r_0$ & $\bar{N}$ \\
    \hline
FJC & [3-20] & [20-80]    & 100.0    & (0.7-1.31]   & ($3\times 10^2$-$2\times 10^4$)
\\
G & [1.5-15] & [16-80]    & [0.867-4.0]    & 0.0   & ($2\times 10^2$-$4\times 10^3$)
\\
Lattice & 2.4 & [16-100]    & $\infty$    & $\sqrt{2}$   & ($9\times 10^2$-$6\times 10^3$)
\\
FTS & 8.0 & [16-64]    & N/A    & N/A   & ($1\times 10^3$-$5\times 10^3$)
\\
\hline
\end{tabular}
\end{table}

\begin{figure}[h!]
\centering
  \begin{subfigure}{0.4\textwidth}
	\includegraphics[width=\linewidth,height=\textheight,keepaspectratio]{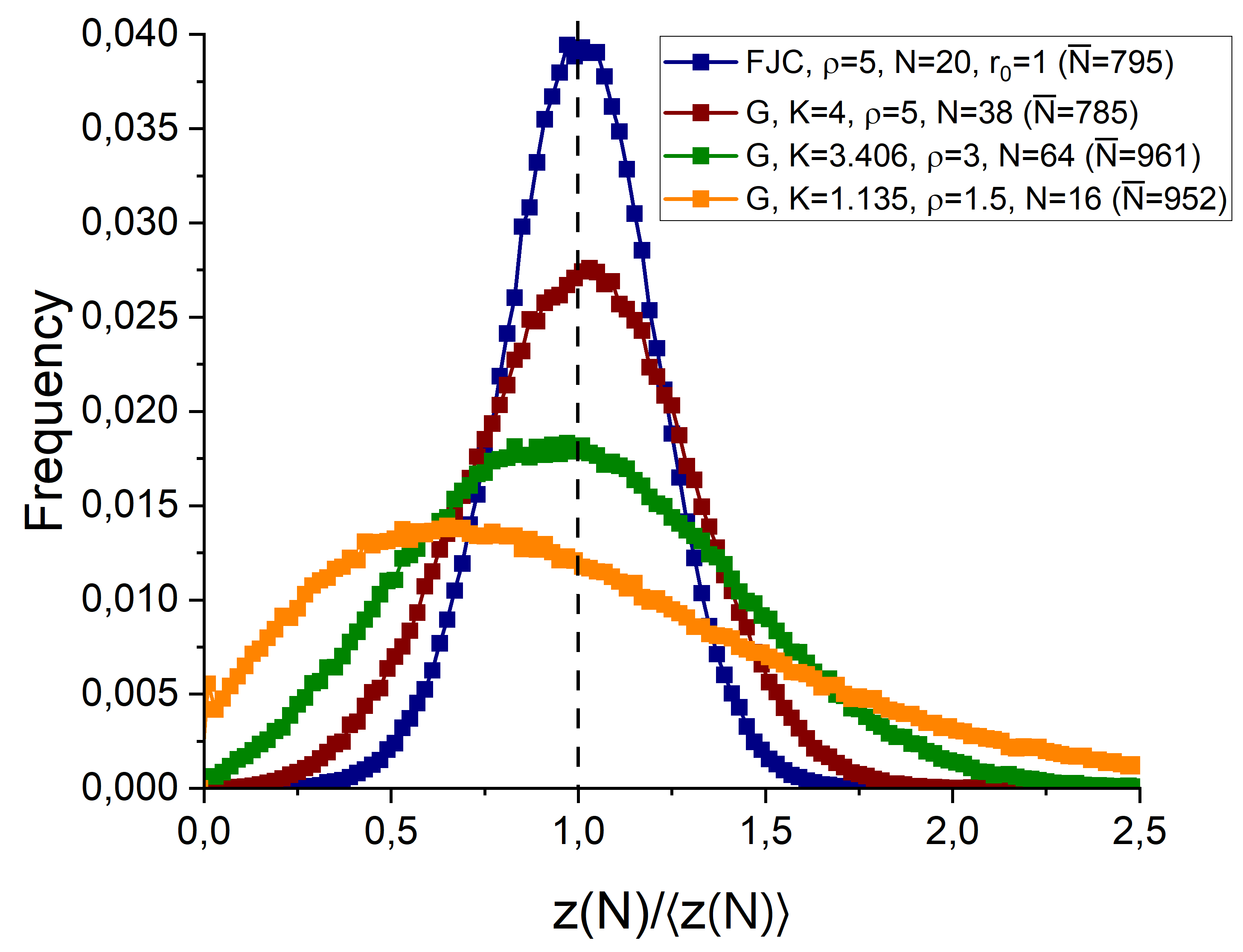}
	\caption{}
	\end{subfigure}
	\begin{subfigure}{0.4\textwidth}
	\includegraphics[width=\linewidth,height=\textheight,keepaspectratio]{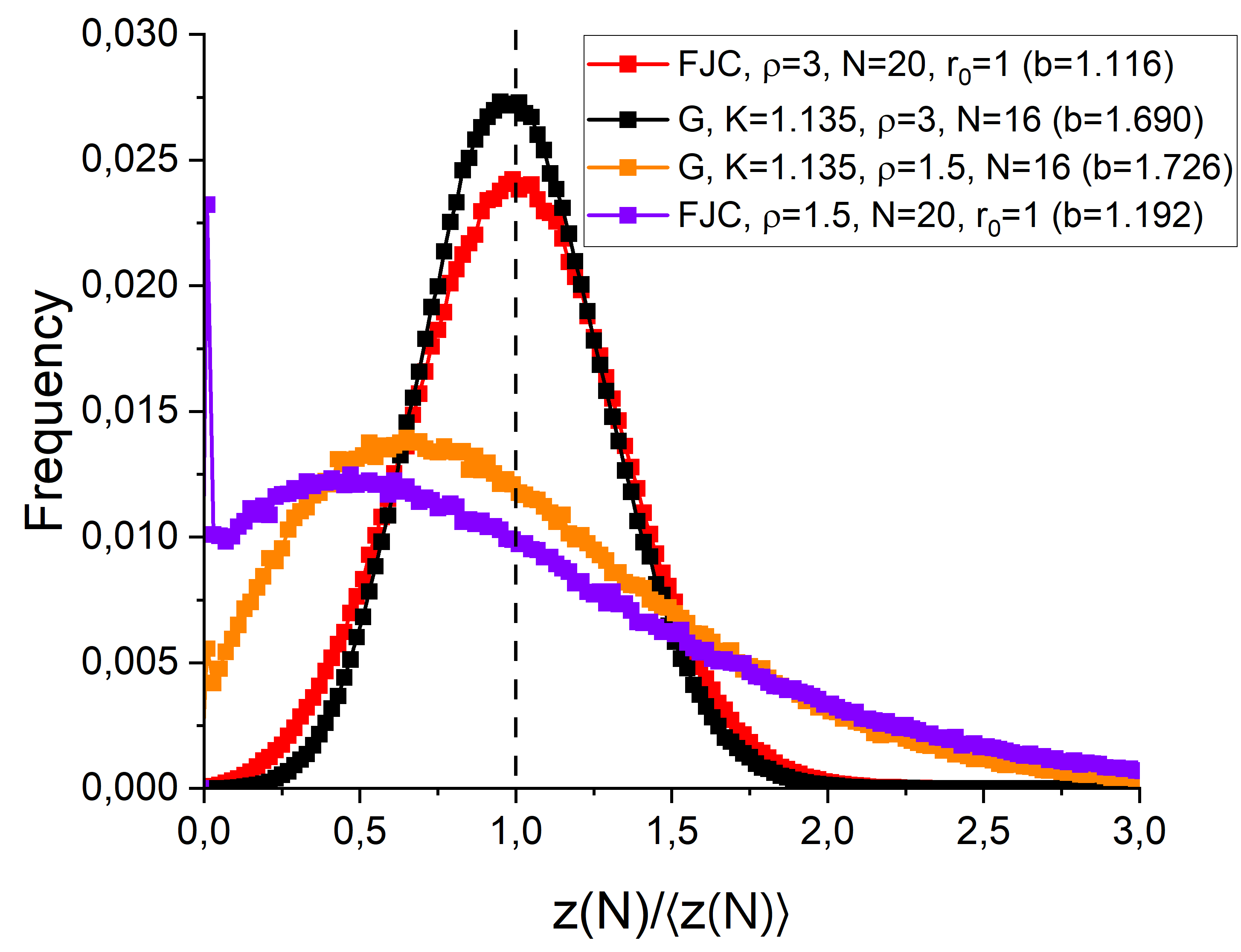}
	\caption{}
	\end{subfigure}
 	\begin{subfigure}{0.4\textwidth}
	\includegraphics[width=\linewidth,height=\textheight,keepaspectratio]{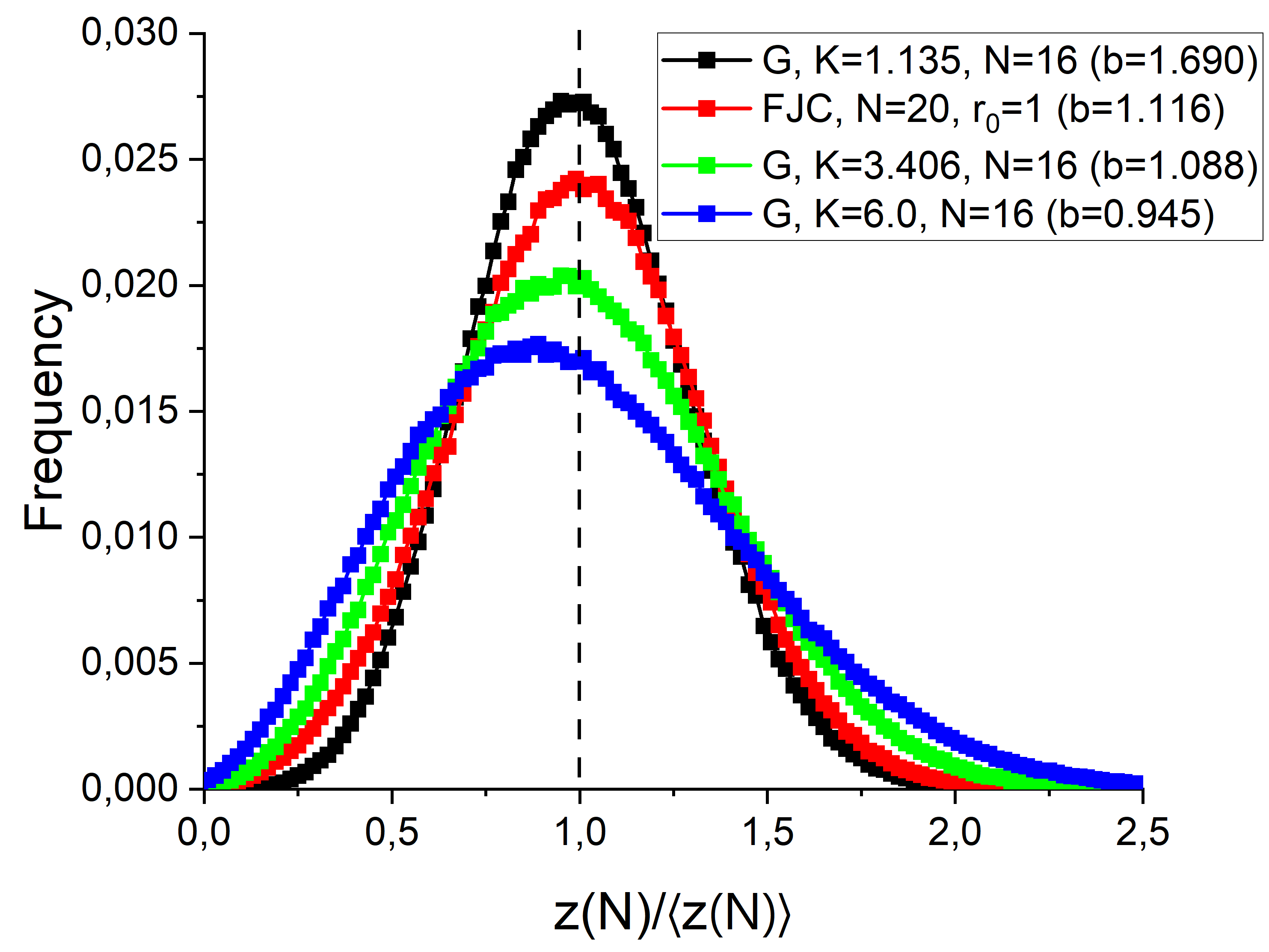}
	\caption{}
	\end{subfigure}
  \caption{The distribution of $z(N)$ in different models of homopolymer melts. The $z(N)$ values were divided by the average $z(N)$ value $\langle z(N)\rangle$ for each model. The vertical dashed line shows $z(N)/\langle z(N)\rangle=1$. (a) Models having similar values of $\bar{N}$. (b) Models having different densities $\rho$. The statistical segment length $b$ for each model is listed in the parentheses. (c) Models having different values of $b$ at the same $\rho=3$.}
  \label{fig:1}
\end{figure}

Before studying the phase transition, we investigated the $z(N)$-distributions of the DPD models at $\chi=0$ (i.e., fully homogeneous phase). For the DPD interaction potential (Eq. \ref{eq:33}), $z(N)$ was calculated as $z(N)=\sum_{i,j}{0.5(1-r_{ij})^2/(NM)}$, where $M$ is the total number of chains in the system, the distance between beads is $r_{ij}<1$, and the beads $i$ and $j$ cannot belong to the same chain. To evaluate whether the parameters of models had any effect on the $z(N)$-distribution, we constructed four different models having roughly similar $\bar{N}$. Fig. \ref{fig:1}a shows that the distributions behaved qualitatively differently depending on the model. For high-density FJC and Gaussian models, the distributions had symmetric, near-Gaussian form. However, upon a decrease of system density below $\rho\approx 3$, the distributions became strongly skewed.

To investigate the effect of the model parameters on the skew of the distributions in more detail, we analyzed the two sets of models having different $\rho$ but similar $b$, $K$, and $N$. Fig. \ref{fig:1}b demonstrates that a decrease of density led to a strong skew of the distributions regardless of the chain model (FJC or Gaussian). Fig. \ref{fig:1}c shows that a decrease of the bond length at moderate density $\rho=3$ also led to a slight skew of the $z(N)$-distribution, although $b$ affected the skew much more weakly than $\rho$ (Fig. \ref{fig:1}b). Finally, making systems more compressible also led to slightly more skewed $z(N)$-distributions (Fig. \ref{fig:s_zn}a).

This behavior can be explained by examining the microscopic structure of a simple DPD liquid. A decrease of $\rho$ leads to a smaller average $z(N)$ and larger relative spread of the $z(N)$-distribution due to a depletion of neighbors around a monomer. Therefore, since $z(N)$ is nonnegative, its distribution becomes right-skewed at low $\rho$ (Fig. \ref{fig:s_zn}b). In a polymer liquid, a bead experiences fewer contacts with monomers from other chains upon a decrease of density or bond length. As a result, the $z(N)$-distribution becomes skewed at some low $\rho$ and $b$ (Fig. \ref{fig:1}b,c). Compressibility of the system also has an effect since less compressible systems have narrower and, therefore, less skewed $z(N)$-distributions (Fig. \ref{fig:s_zn}a). However, despite an abundance of parameters defining the shape of the $z(N)$-distribution, the system density plays the dominant role (Fig. \ref{fig:1}b).

Next, we studied how the $z(N)$-distribution affected the behavior of systems at ODT. We established the ODT point location $(\chi N)_\text{ODT}(\bar{N})$ for all studied models (Table \ref{table}, \ref{table2}). We assumed the linear relation between $\chi$ and $\alpha$ ($\chi=z\alpha/k_BT$). Following Morse and co-workers,\cite{glaser2014universality} the value of $z$ was calculated for each model as the $N\to \infty$ limit of $z(N)$ measured at $\chi=0$ (see SI, section 2) . To determine $(\chi N)_\text{ODT}$ as $(\chi N)_\text{ODT}=zN\alpha_\text{ODT}/k_BT$, we established $\alpha_\text{ODT}$ as the average between the lowest and highest values of $\alpha$ leading to ordering and disordering, respectively (see SI section 3).

\begin{figure}[h!]
\centering
  \begin{subfigure}{0.46\textwidth}\includegraphics[width=\linewidth,height=\textheight,keepaspectratio]{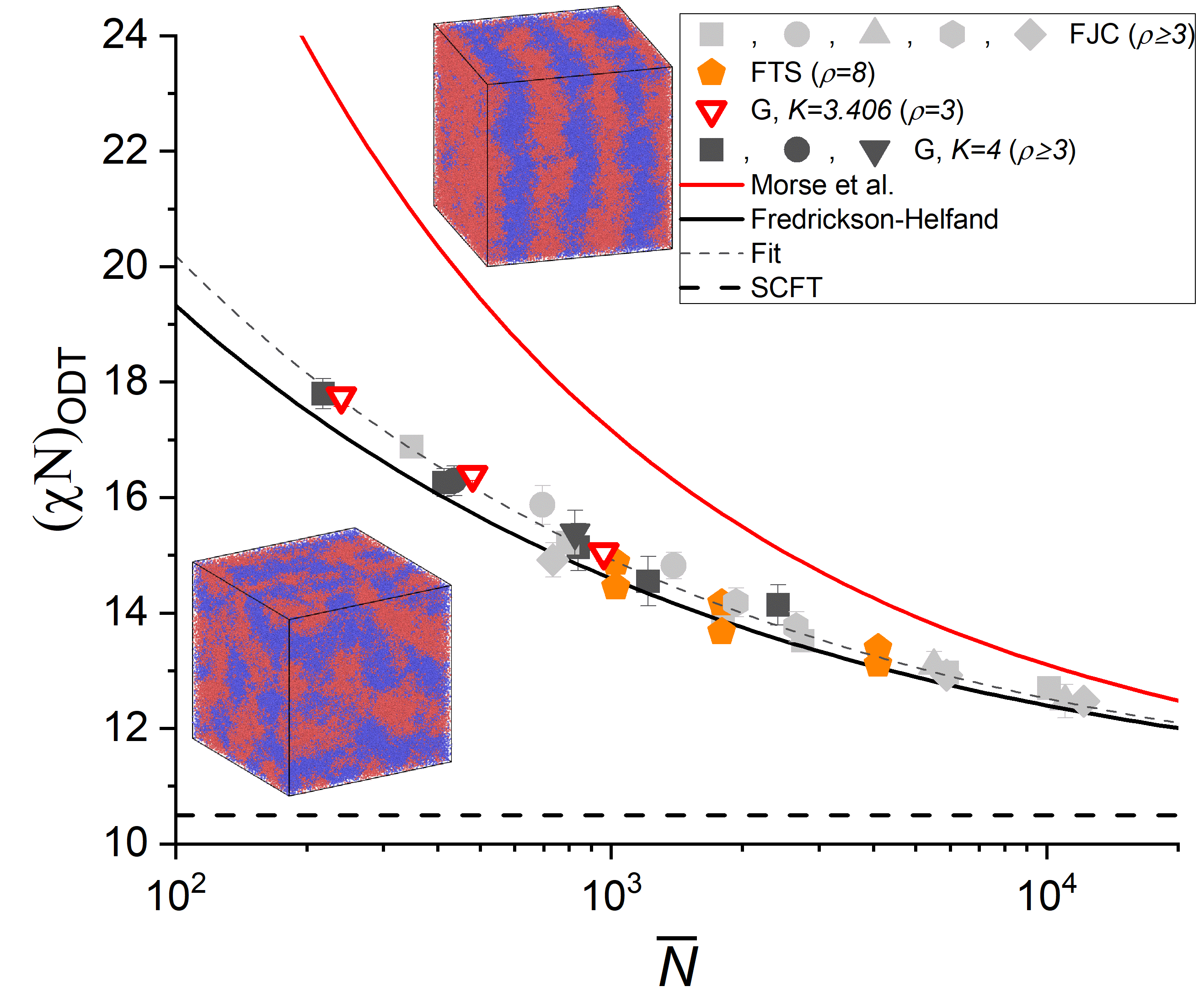}
	\caption{}
	\end{subfigure}
	\begin{subfigure}{0.49\textwidth}\includegraphics[width=\linewidth,height=\textheight,keepaspectratio]{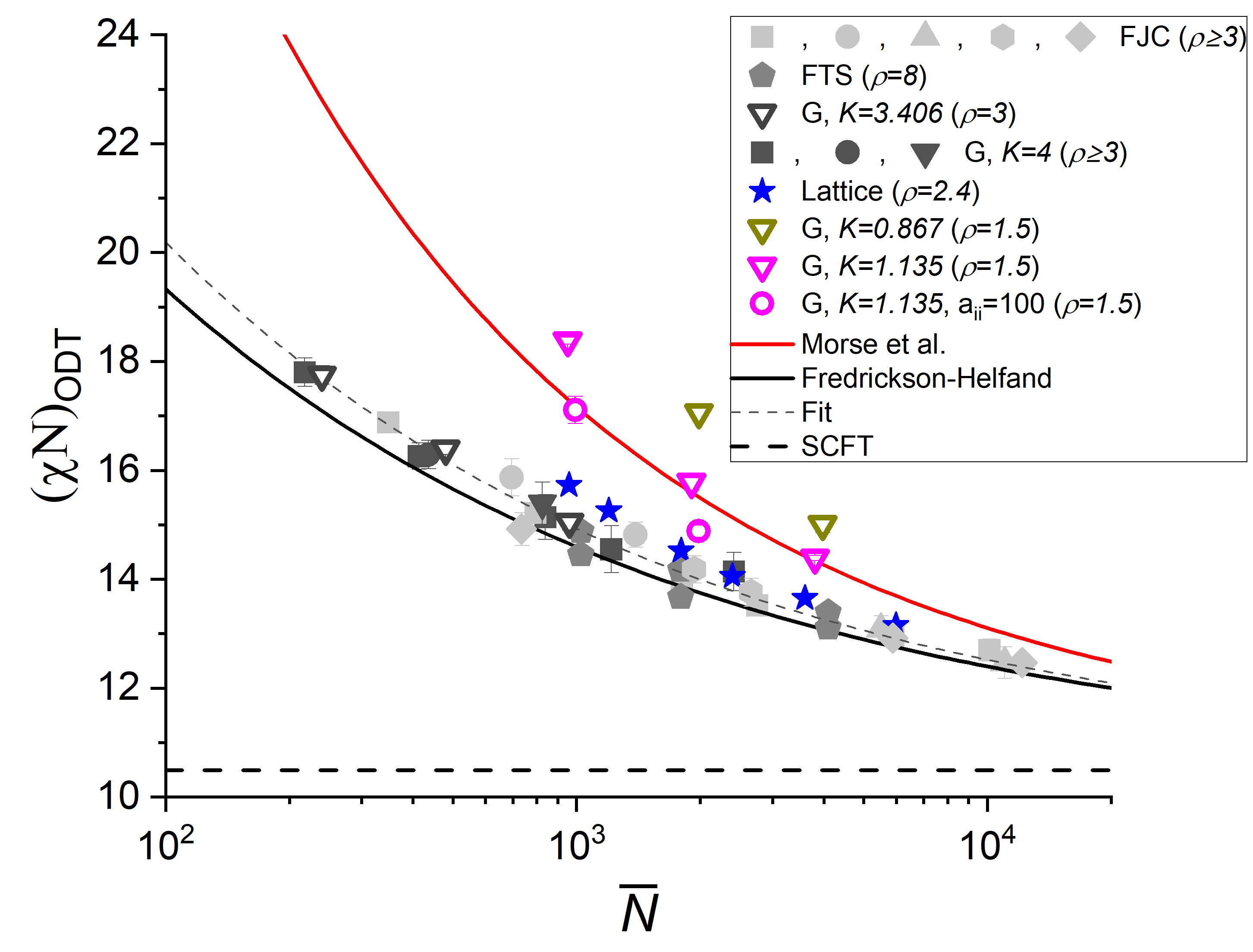}
	\caption{}
	\end{subfigure}
  \caption{$(\chi N)_\text{ODT}(\bar{N})$ dependencies. Red, black, and thick dashed curves represent the predictions of Eqs. \ref{eq:2}, \ref{eq:1}, and SCFT, respectively. (a) The data for the models having symmetric $z(N)$-distribution.
  Snapshots show the disordered (bottom) and ordered (top) states near ODT for the FJC model with $\rho=3$, $N=80$. (b) The data for all studied models. The models having symmetric $z(N)$-distribution (data from (a)) are shown in gray, colored symbols represent the models with skewed $z(N)$-distributions. Table \ref{table2} contains correspondence between the symbol shape and the set of model parameters.}
  \label{fig:2}
\end{figure}

\begin{figure}[h!]
\centering
  \begin{subfigure}{0.49\textwidth}\includegraphics[width=\linewidth,height=\textheight,keepaspectratio]{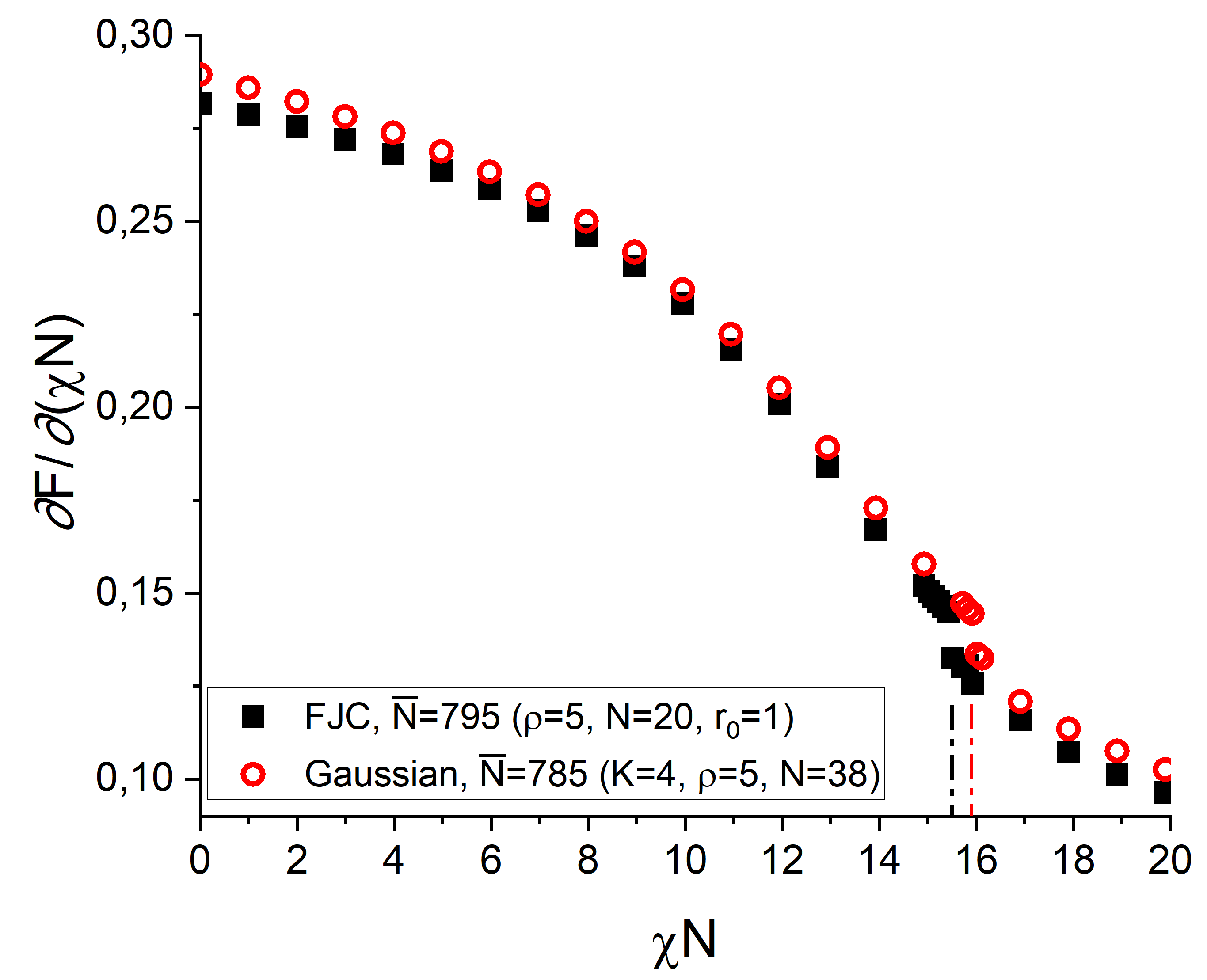}
	\caption{}
	\end{subfigure}
	\begin{subfigure}{0.49\textwidth}\includegraphics[width=\linewidth,height=\textheight,keepaspectratio]{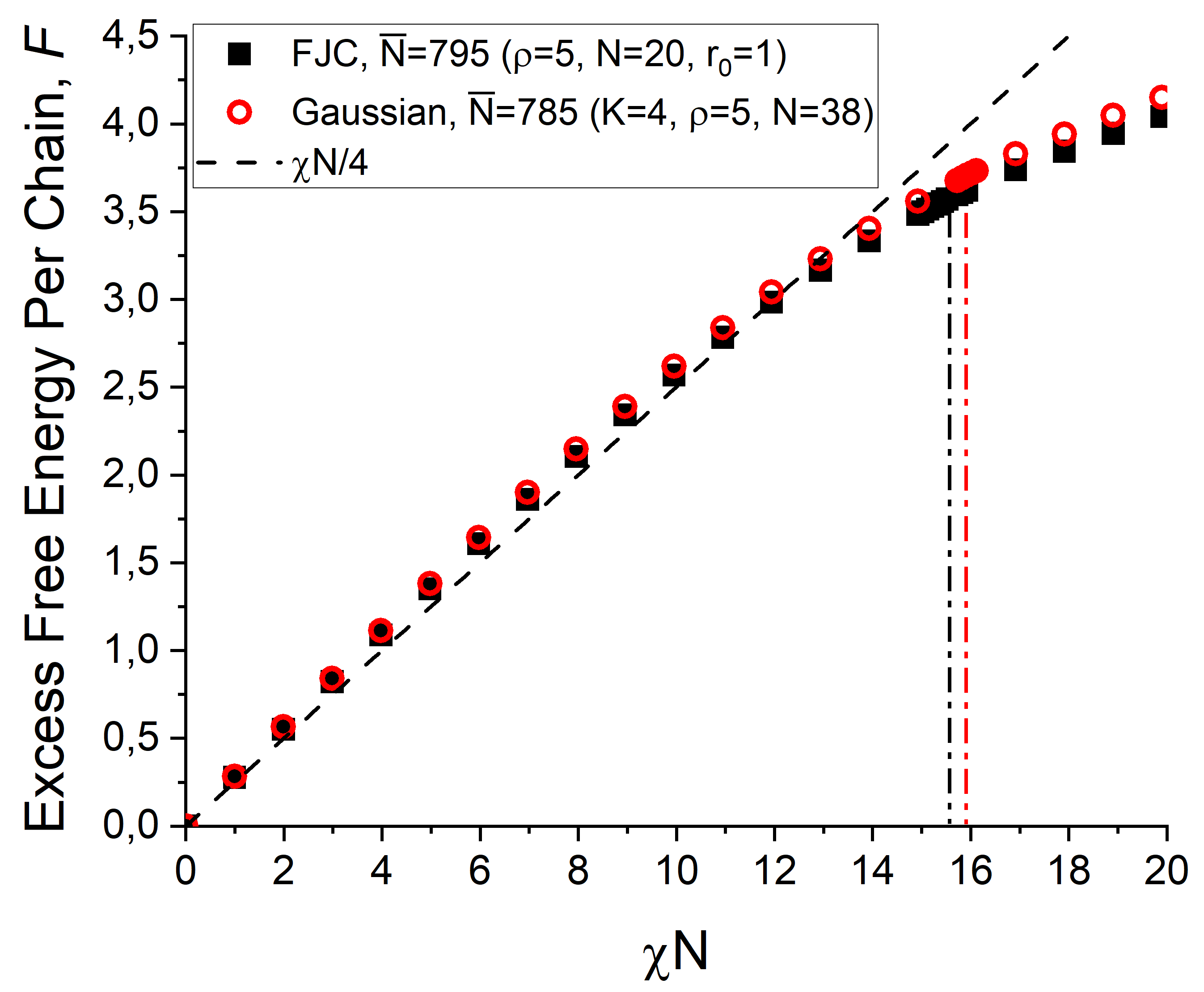}
	\caption{}
	\end{subfigure}
 \caption{(a) The $\chi N$-dependencies of the first derivative of the excess free energy per chain for an FJC model (black dots) and for a Gaussian model with $K=4$ (red circles) having similar $\bar{N}\approx 800$. The derivative was estimated using the average potential energy of A-B interaction as outlined in ref. \cite{glaser2014universality}. Both dependencies exhibited a discontinuity at ODT: vertical dash-dotted lines represent the lowest possible $\chi N$ values at which ordering occurs for the FJC model (black) and for the Gaussian model (red). (b) The dependencies of the excess free energy per chain in the two models calculated by numerically integrating (a). The black dashed line shows the mean-field dependency in the disordered phase ($F=\chi N/4$).}
  \label{fig:3}
\end{figure}

Our main results are presented in Fig. \ref{fig:2}. Fig. \ref{fig:2}a shows the $(\chi N)_\text{ODT}(\bar{N})$ dependency for all models having nearly-symmetric distribution of $z(N)$. We included the field-theoretic simulation (FTS) data in Fig. \ref{fig:2}a, since ref. \cite{alexander2007diblock} demonstrated that the microscopic field distribution, proportional to $z(N)$, is symmetric in FTS at the studied values of $\bar{N}$. Fig. \ref{fig:2}a shows that all $(\chi N)_\text{ODT}$ values obtained at similar invariant chain length $\bar{N}$ overlapped within the error of $(\chi N)_\text{ODT}$ determination, demonstrating the universal behavior of the models. The excess chain free energy and its first derivative agreed well at similar $\bar{N}$, thus confirming the universality (Fig. \ref{fig:3}). Moreover, fitting of the DPD simulation data in Fig. \ref{fig:2}a by the expression $(\chi N)_\text{ODT}=10.495+A\bar{N}^{-B}$ yielded $A=46.4\pm 2.3$ and $B=0.340\pm 0.008$ (thin gray dashed curve), which agrees well with the prediction of the FH theory (Eq. \ref{eq:1}) within the error.

Fig. \ref{fig:2}b shows the $(\chi N)_\text{ODT}(\bar{N})$ dependencies for all studied models. The models having skewed $z(N)$-distributions showed nonuniversal behavior in the $\chi\propto\alpha$ approximation and did not follow Eq. \ref{eq:1} or \ref{eq:2}. We investigated how the $z(N)$-distribution skew affected the deviation from the universality by making systems four times less compressible (the repulsion parameter between like-type beads $a_{ii}$ was increased from $a_{ii}=25$ to $a_{ii}=100$ for the Gaussian model with $K=1.135$). Fig. \ref{fig:s_zn}a shows that less compressible systems had more symmetric $z(N)$-distributions; Fig. \ref{fig:2}b shows that such models had $(\chi N)_\text{ODT}$ points lying closer to the FH line (Eq. \ref{eq:1}). Therefore, a decrease of the $z(N)$-distribution skew can drive models to the universality in the $\chi\propto\alpha$ approximation. Finally, the multiple-occupancy lattice models with the average polymer density $\rho=2.4$ adhered to the FH universality at high $\bar{N}$ but started to deviate from this trend at $\bar{N}\lesssim 10^3$. This deviation was observed presumably due to the density small enough to cause a significant skew of the $z(N)$-distribution.

The distribution of $z(N)$ characterizes the microscopic fluctuations of the potential energy of a monomer in a melt. Especially for low-density models, the $z(N)$-distribution becomes skewed and leads to a poorly-defined average $\langle z \rangle$ that no longer coincides with the most probable $z$. Therefore, the $\chi=z\alpha/k_BT$ approximation becomes inapplicable to such models and leads to nonuniversal behavior. Fig. \ref{fig:s_median} shows Fig. \ref{fig:2}b in which $z$ was defined as the $N\to\infty$ limit of the \textit{median} values of $z(N)$; it shows agreement with FH theory within the uncertainty associated with the difference between the mean, median, and the mode of the skewed $z(N)$-distributions.
Apparently, for the models with skewed $z(N)$-distributions, the change of the local liquid structure with $\alpha$ becomes so significant that the universality can be recovered only by adding nonlinear terms to the $\chi(\alpha)$ function; these terms can be calculated solely from the \textit{mesoscale} structure of the melt (peak of the structure factor) at different $\alpha$ as opposed to the \textit{microscopically} determined homopolymer melt characteristic $z$ defining the linear part of $\chi(\alpha)$ \cite{glaser2014universality}.
The nature of those nonlinear terms is unclear; however, they help to describe strong fluctuations at high $\alpha$ that are treated incorrectly in the $\chi\propto\alpha$ approximation \cite{qin2012fluctuations} and, we believe, "account" for the $z(N)$-distribution skew.

To conclude, we found that models of block copolymer melts exhibit universal behavior in the linear $\chi \propto \alpha $  approximation if the distribution of the monomer interaction potential energy $z(N)$ is symmetric. In addition, the models with symmetric $z(N)$-distributions yielded the universal $(\chi N)_\text{ODT}(\bar{N})$ dependency agreeing well with the FH scaling (Eq. \ref{eq:1}) and demonstrated that the FH theory can correctly predict the ODT point well beyond its assumed validity range of $\bar{N}>10^4$ under the microscopic restriction we mentioned above.  This simple definition of $\chi$ and the universality of the $(\chi N)_\text{ODT}(\bar{N})$ function in the experimentally relevant range $\bar{N}>10^2$ will allow researchers to compare the results of simulations and experiments without performing the complex and resource-demanding Morse calibration. In addition, we expect that block copolymers of various architectures will behave universally in the $\chi\propto\alpha$ approximation if the $z(N)$-distribution is symmetric; this is the matter of our next study.
As a final note, we would like to point out that FJC models having symmetric $z(N)$-distributions are particularly suitable for reverse mapping onto atomistic models of real polymer systems, since one can map a Kuhn segment to a bead of an FJC model, and the resulting model will have typical densities high enough to yield a symmetric $z(N)$-distribution (see details in SI section 6). This might be quite beneficial for experimental and computational polymer scientists interested in block copolymer mesostructure design and polymer phase behavior.

We thank Gabriella La Cour, Tianyi Jin, YongJoo Kim, Jian Qin, Mark Matsen, and Alexey Gavrilov for help and fruitful discussions. We acknowledge the MIT SuperCloud and Lincoln Laboratory Supercomputing Center for providing computational resources. This research was supported by the National Science Foundation through award number DMREF 2118678.

\section{Supporting Information}
\subsection{Simulation Details}
In this section, we will describe the details of the dissipative particle dynamics (DPD) simulation carried out in this work. We used DPD implementation in LAMMPS software. The bead-and-spring model was adopted to simulate polymers. Conservative forces $\bold{F}_{ij}^c$, random forces, and dissipative forces acted on all beads. The conservative force modeled soft repulsion between beads and was determined by Eq. \ref{eq:33}.

\begin{equation}
\label{eq:33}
\bold{F}_{ij}^c=
\begin{cases}
a_{xy}\left(1-r_{ij}\right)\frac{\bold{r}_{ij}}{r_{ij}}, \quad r_{ij}\leq 1, \\
0, \quad r_{ij} > 1.
\end{cases}
\end{equation}
In Eq. \ref{eq:33}, $a_{xy}$ denotes the repulsion parameter between beads $i$ and $j$ having types $x$ and $y$, respectively ($x=A,B$, $y=A,B$). In these notations, $\alpha=a_{AB}-a_{xx}$, since $a_{xx}=a_{AA}=a_{BB}$. We calculated $a_{xx}$ as $a_{xx}=75k_BT/\rho$, where $\rho$ is the number of beads per unit volume, to enforce compressibility of the system similar to the compressibility of water \cite{groot1997dissipative}. The only exception were the Gaussian chain models at density $\rho=1.5$; in these models, $a_{xx}$ was equal to $a_{xx}=25$ unless stated otherwise. For a given polymer model, we varied $a_{AB}$ to vary $\alpha$. 

We carried out simulations in NVT ensemble, the temperature was kept equal to $k_BT=1$. We utilized fixed-length cubic boxes with periodic boundary conditions applied in all directions, the box side length $L$ was adjusted depending on the system to ensure that there were at least $\#_{Lam}=3$ full lamellar periods in the box at the ODT point (Table \ref{table2}). In fact, we had $\#_{Lam}\geq 4$ in the majority of the systems; therefore, the finite box size effects that are often associated with NVT simulations of ODT \cite{medapuram2015universal,gavrilov2023effect} were negligible (Supporting Information, section 4). The integration time step was equal to $\Delta t=0.015$, the maximal simulation time was set to $3\times 10^7$ steps. Other simulation parameters were set as discussed elsewhere \cite{groot1997dissipative}.

\newpage
\begin{table}
 \caption{Parameters of the simulated models. The rightmost column contains description of the symbol used for each model in Fig. 2b in the main text. The last two models were simulated with $a_{xx}$ parameter equal to $a_{xx}=100$.}
  \begin{tabular}{@{}lllllllllll@{}}
    \hline
    $\rho$ & $N$ & $K$ & $r_0$ & $L$ & $\#$ of chains & $\#_{Lam}$ & $z$ & $\alpha_\text{ODT}/k_BT$ & $\bar{N}$ & Fig. 2b symbol\\
    \hline
3 & 20    & 100    & 1.0   & 30  & 4050 & 5 & 0.2837 & 2.9750 & 348 & pale gray square
\\
3 & 40    & 100    & 1.0   & 37.8  & 4050 & 4 & 0.2837 & 1.3983 & 696 & pale gray circle
\\
3 & 80    & 100    & 1.0 & 47.6 & 4044 & 3 & 0.2837 & 0.6528 & 1393 & pale gray circle
\\
5 & 20    & 100     & 1.0   & 30  & 6750 & 5 & 0.6607 & 1.1520 & 795 & pale gray square
\\
5 & 20     & 100    & 1.18   & 37.8  & 13502 & 6 & 0.7024 & 1.0095 & 1935 & pale gray hexagon
\\
5 & 20    & 100     & 1.26   & 37.8  & 13502 & 5 & 0.7120 & 0.9664 & 2657 & pale gray hexagon
\\
8 & 20     & 100    & 1.0   & 30  & 10800 & 5 & 1.2612 & 0.5531 & 1805 & pale gray square
\\
10 & 20     & 100    & 1.0   & 30  & 13500 & 5 & 1.6707 & 0.4047 & 2752 & pale gray square
\\
10 & 40     & 100    & 1.0   & 37.8  & 13502 & 5 & 1.6707 & 0.1959 & 5504  & pale gray triangle
\\
10 & 80     & 100    & 1.0   & 37.8  & 6751 & 4 & 1.6707 & 0.0933 & 11007  & pale gray triangle
\\
10 & 20     & 100    & 0.76745   & 30  & 13500 & 6 & 1.5089 & 0.4945 & 735 & pale gray diamond
\\
10 & 20      & 100   & 1.14873   & 30  & 13500 & 5 & 1.7167 & 0.3762 & 5885 & pale gray diamond
\\
10 & 20      & 100   & 1.31   & 30  & 13500 & 4 & 1.7433 & 0.3577 & 12161 & pale gray diamond
\\
15 & 20      & 100   & 1.0   & 30  & 20250 & 5 & 2.7040 & 0.2398 & 5907 & pale gray square
\\
20 & 20      & 100   & 1.0   & 30  & 27000 & 5 & 3.7443 & 0.1696 & 10116 & pale gray square
\\
3 & 20    & 4.0    & 0.0   & 30.0  & 4050 & 5 & 0.2193 & 4.0545 & 218 & dark gray square
\\
3 & 40    & 4.0    & 0.0 & 37.8 & 4050 & 4 & 0.2193 & 1.8573 & 436 & dark gray circle
\\
5 & 20    & 4.0     & 0.0   & 30.0  & 6750 & 5 & 0.5479 & 1.4868 & 413 & dark gray square
\\
5 & 40     & 4.0    & 0.0   & 37.8  & 6751 & 5 & 0.5479 & 0.702 & 826 & dark gray triangle
\\
8 & 20    & 4.0     & 0.0   & 30.0  & 10800 & 6 & 1.1090 & 0.6825 & 840 & dark gray square
\\
10 & 20     & 4.0    & 0.0   & 30.0  & 13500 & 6 & 1.5014 & 0.4845 & 1214 & dark gray square
\\
15 & 20     & 4.0    & 0.0   & 30.0  & 20250 & 7 & 2.5099 & 0.2818 & 2416 & dark gray square
\\
3 & 16    & 3.406    & 0.0   & 30.0  & 5062 & 4 & 0.2363 & 4.6962 & 241 & gray empty triangle
\\
3 & 32    & 3.406    & 0.0 & 30.0 & 2531 & 3 & 0.2363 & 2.1675 & 482 & gray empty triangle
\\
3 & 64    & 3.406     & 0.0   & 37.8  & 2531 & 3 & 0.2363 & 0.9951 & 964 & gray empty triangle
\\
1.5 & 16     & 1.135    & 0.0   & 30.0  & 2531 & 3 & 0.09195 & 12.5053 & 952 & magenta triangle
\\
1.5 & 32    & 1.135     & 0.0   & 37.8  & 2531 & 3 & 0.09195 & 5.3698 & 1904 & magenta triangle
\\
1.5 & 64     & 1.135    & 0.0   & 47.6  & 2527 & 3 & 0.09195 & 2.4477 & 3808 & magenta triangle
\\
1.5 & 16    & 0.867    & 0.0 & 30.0 & 2531 & 3 & 0.09814 & 10.852 & 1988 & dark yellow triangle
\\
1.5 & 32    & 0.867     & 0.0   & 37.8  & 2531 & 3 & 0.09814 & 4.7756 & 3976 & dark yellow triangle
\\
1.5  & 16     & 1.135    & 0.0   & 30.0  & 2531 & 3 & 0.04298 & 24.885 & 991 & magenta circle
\\
1.5  & 32    & 1.135    & 0.0   & 37.8  & 2531 & 3 & 0.04298 & 10.861 & 1982 & magenta circle
\\
    \hline
  \end{tabular}
  \label{table2}
\end{table}

\clearpage
\subsection{Determination of the $b$ and $z$ parameters}
In this section, we describe the calculation of model-dependent parameters $b$ and $z$ which determine the values of $\bar{N}$ and $\chi$. To establish the value of $b$, we followed the method outlined in ref. \cite{glaser2014universality}. First, we fixed the values of all model parameters and simulated three homopolymer melt systems (i.e., at $\alpha=0$) with chains of length $N=20$, $N=40$, and $N=80$, respectively, (or, alternatively, having length $N=16$, $N=32$, and $N=64$ for the Gaussian models with $K\leq 3.406$, see Table \ref{table2}) for $2\times 10^6$ DPD time steps. Second, we calculated the squared radius of gyration of chains $R_g^2(N)$; for each system, we averaged the $R_g^2$ values across 10 structures obtained every $10^5$ steps during the last $10^6$ DPD time steps of the simulation. Finally, we obtained $b$ by fitting the $R_g^2(N)$ dependency by the ROL-theoretic expression in Eq. \ref{eq:4}, where $b$ and $\gamma$ are fitting parameters.

\begin{equation}
\label{eq:4}
    \frac{6R^2_g}{N}=b^2\left(1-\frac{1.42}{\rho b^3N^{1/2}}+\frac{\gamma}{N}\right)
\end{equation}

To determine $z$, we first measured the effective coordination number $z(N)$ defined for the DPD interaction potential (Eq. \ref{eq:33}) as $z(N)=\sum_{i,j}{0.5(1-r_{ij})^2/(NM)}$, where $M$ is the total number of chains in the system, the distance between beads is $r_{ij}<1$, and the beads $i$ and $j$ cannot belong to the same chain. For the given set of model parameters, we calculated $z(N)$ for each of the three aforementioned homopolymer melt systems and averaged those values as described in the previous paragraph. Second, using the calculated value of $b$, the fit of the $z(N)$ dependency by the formula in Eq. \ref{eq:5} yields $z$ and $\delta$ as fitting parameters \cite{morse2009chain,glaser2014universality}.

\begin{equation}
\label{eq:5}
    z(N)=z\left(1+\frac{(6/\pi)^{3/2}}{\rho b^3N^{1/2}}+\frac{\delta}{N}\right)
\end{equation}

\subsection{Determination of the ODT Point}

We determined the $\chi_\text{ODT}$ value according to the simple procedure following the work by Gavrilov \cite{gavrilov2023effect}. We calculated $\chi_\text{ODT}$ as $\chi_\text{ODT}$=$(\chi_{or}+\chi_{dis})/2$, where $\chi_{or}$ is the lowest value of $\chi$ at which ordering from a homogeneous diblock copolymer melt was observed. $\chi_{dis}$ is the highest value of $\chi$ that yielded disordering of an initially ordered lamellar structure. The difference between $\chi_{or}$ and $\chi_{dis}$ gave the error of $(\chi N)_\text{ODT}$ determination (Fig. 2 in the main text). It is worth mentioning that in that Figure, the error bars for Gaussian models with $K=4$ are only estimates; for those models, we did not follow the procedure of determining the smallest possible error of $\chi_\text{ODT}$ described below because modeling high-density systems with fluctuating bonds imposes strong computational restrictions due to large neighbor lists.

We determined whether systems were in ordered or disordered state by visual analysis; the structures maintained the disordered state after spontaneous disordering and ordered state after ordering for at least $10^6$ DPD time steps. We confirmed that $\chi_{or}$ is indeed the lowest and $\chi_{dis}$ is the highest possible values of $\chi$ leading to ordering and disordering, respectively, as follows. We performed the simulations of a disordered melt at $\chi N=\chi_{or} N-0.1$ for $3\times 10^7$ DPD time steps. If the system stayed disordered, we confirmed that $\chi_{or}$ is indeed the lowest value of $\chi$ that yields spontaneous ordering. Similarly, to confirm the value of $\chi_{dis}$, we carried out the simulations of the lamellar phase at $\chi N=\chi_{dis} N+0.1$. If the system stayed ordered for $3\times 10^7$ DPD time steps, we established that $\chi_{dis}$ is the highest possible value of $\chi$ at which the lamellar structure spontaneously disorders.

\begin{figure}[h!]
\centering
  \begin{subfigure}{0.4\textwidth}
	\includegraphics[width=\linewidth,height=\textheight,keepaspectratio]{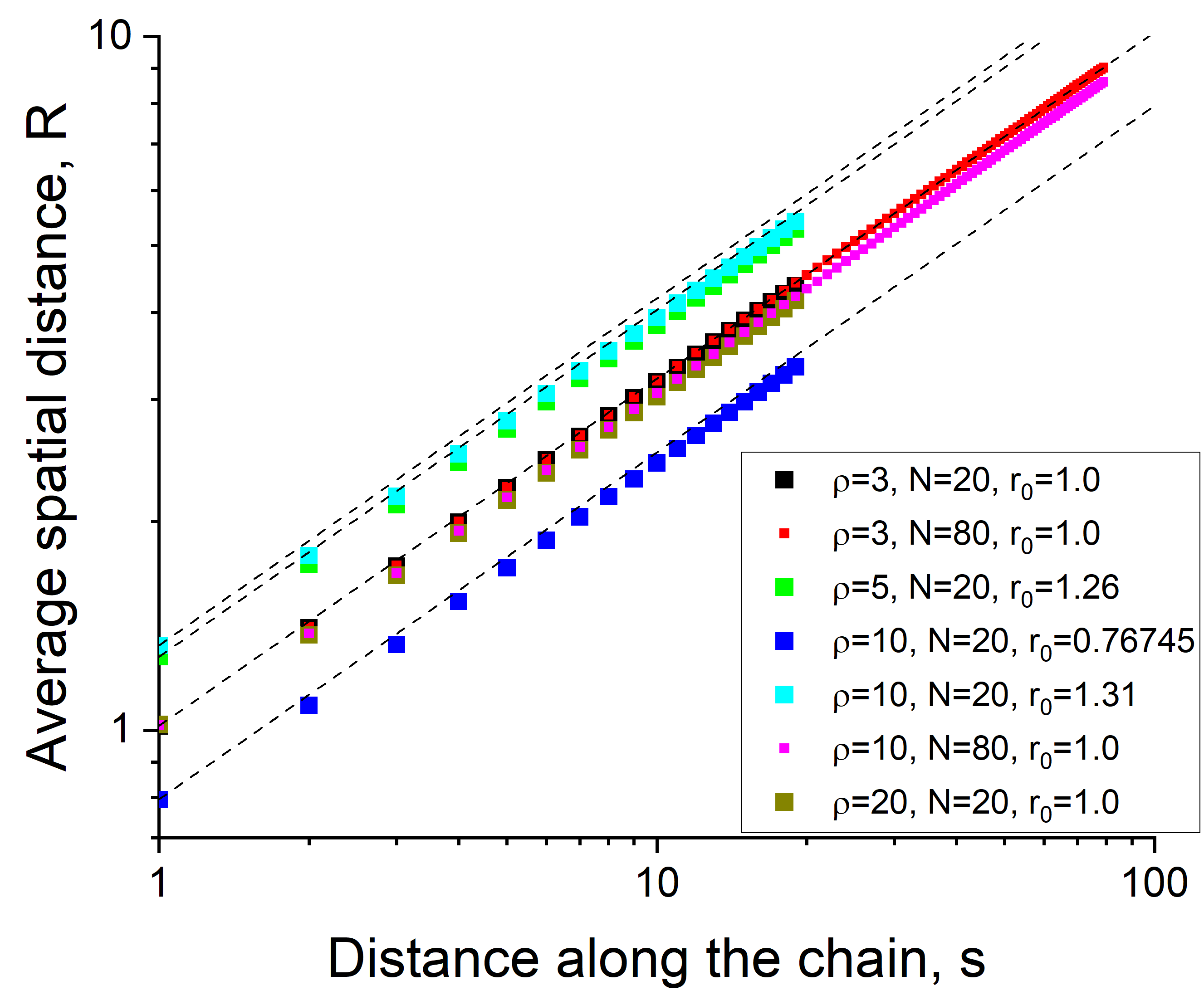}
	\caption{}
	\end{subfigure}
	\begin{subfigure}{0.4\textwidth}
	\includegraphics[width=\linewidth,height=\textheight,keepaspectratio]{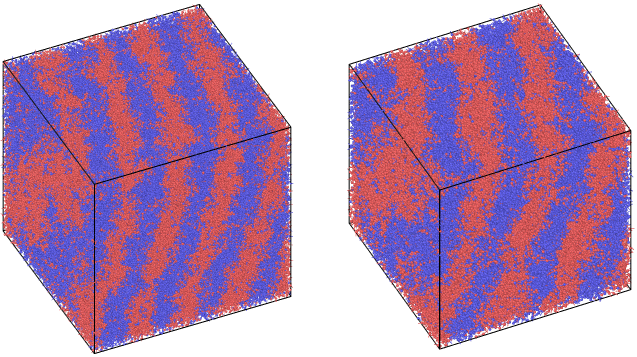}
	\caption{}
	\end{subfigure}
  \caption{(a) The dependencies of the average spatial distance between beads separated by $s$ beads along the chain for several FJC models of melts at $\alpha=0$. The black dashed lines represent the ideal FJC scaling $R=r_{eq}s^{1/2}$, where $r_{eq}=R(1)$ is the equilibrium bond length. (b) The snapshots of the ordered lamellar structures used as initial structures for the simulations of disordering in the FJC system with $\rho=5$, $r_0=1.26$ (see Table \ref{table2}). Red and blue beads represent A and B beads, respectively. Left - box size $L=37.8$, right - $L=30$. For the larger system, $\chi N$ at which this ordered structure was obtained equals to $\chi N=14.1$, for the smaller one $\chi N=14.5$.}
  \label{fig:s1}
\end{figure}

\begin{figure}[h!]
\centering
  \begin{subfigure}{0.49\textwidth}
	\includegraphics[width=\linewidth,height=\textheight,keepaspectratio]{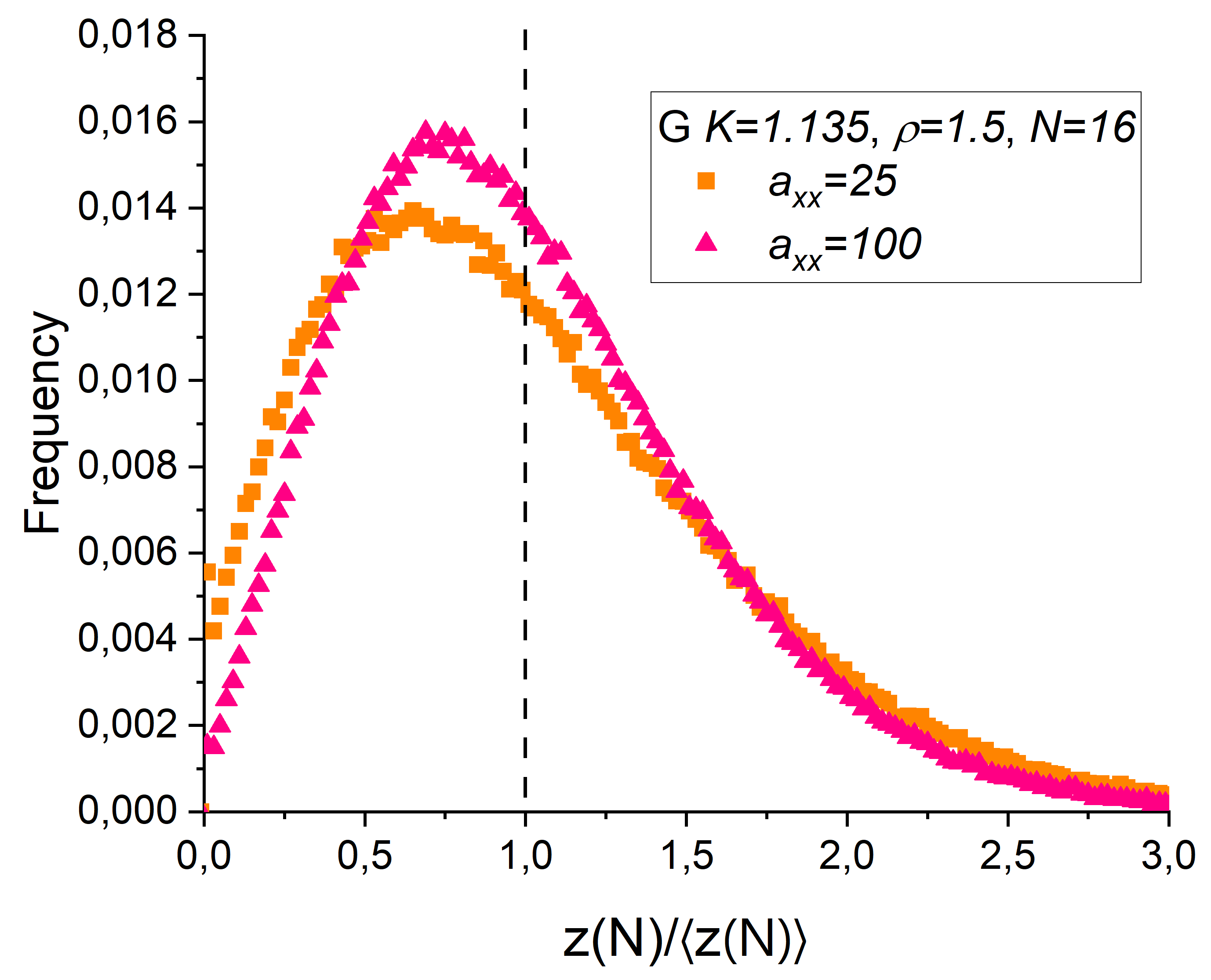}
	\caption{}
	\end{subfigure}
	\begin{subfigure}{0.49\textwidth}
	\includegraphics[width=\linewidth,height=\textheight,keepaspectratio]{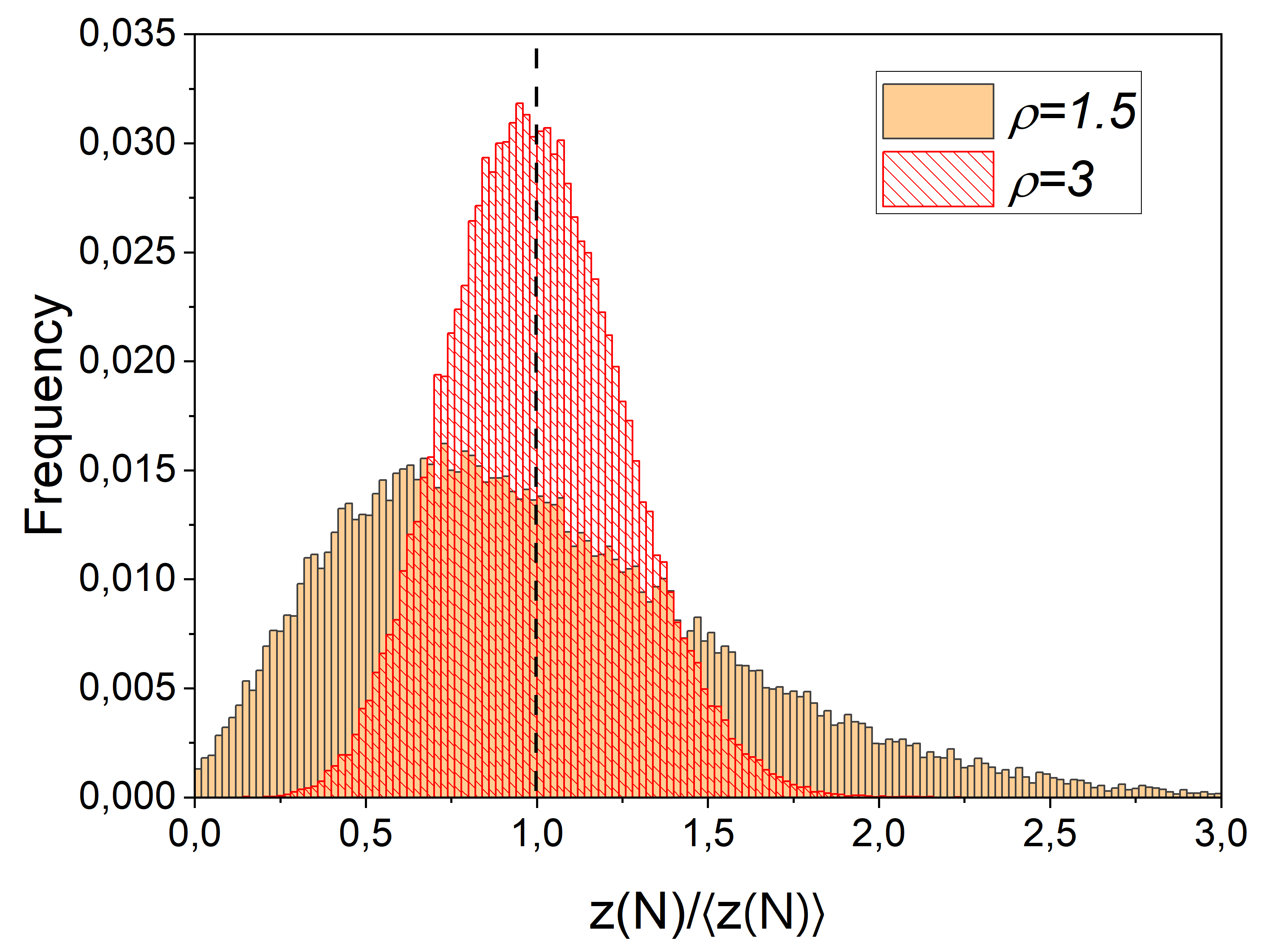}
	\caption{}
	\end{subfigure}
  \caption{The distributions of $z(N)$ values divided by the average $z(N)$ value $\langle z(N)\rangle$ for each model. The vertical dashed line shows $z(N)/\langle z(N)\rangle=1$. (a) $z(N)$-distributions for the Gaussian model with $K=1.135$ at $\alpha=0$. Incompressibility of the system grows with the repulsion coefficient between like-type beads $a_{xx}$ at fixed $\rho$ (Supporting Information, section 1). More incompressible system ($a_{xx}=100$) has a narrower and less skewed distribution. (b) $z(N)$-distributions in two single-phase DPD liquids ($\alpha=0$, $N=1$, no bonds) at density $\rho=1.5$ (yellow) and $\rho=3$ (red).}
  \label{fig:s_zn}
\end{figure}

\begin{figure}[h!]
\centering
  \includegraphics[width=\linewidth]{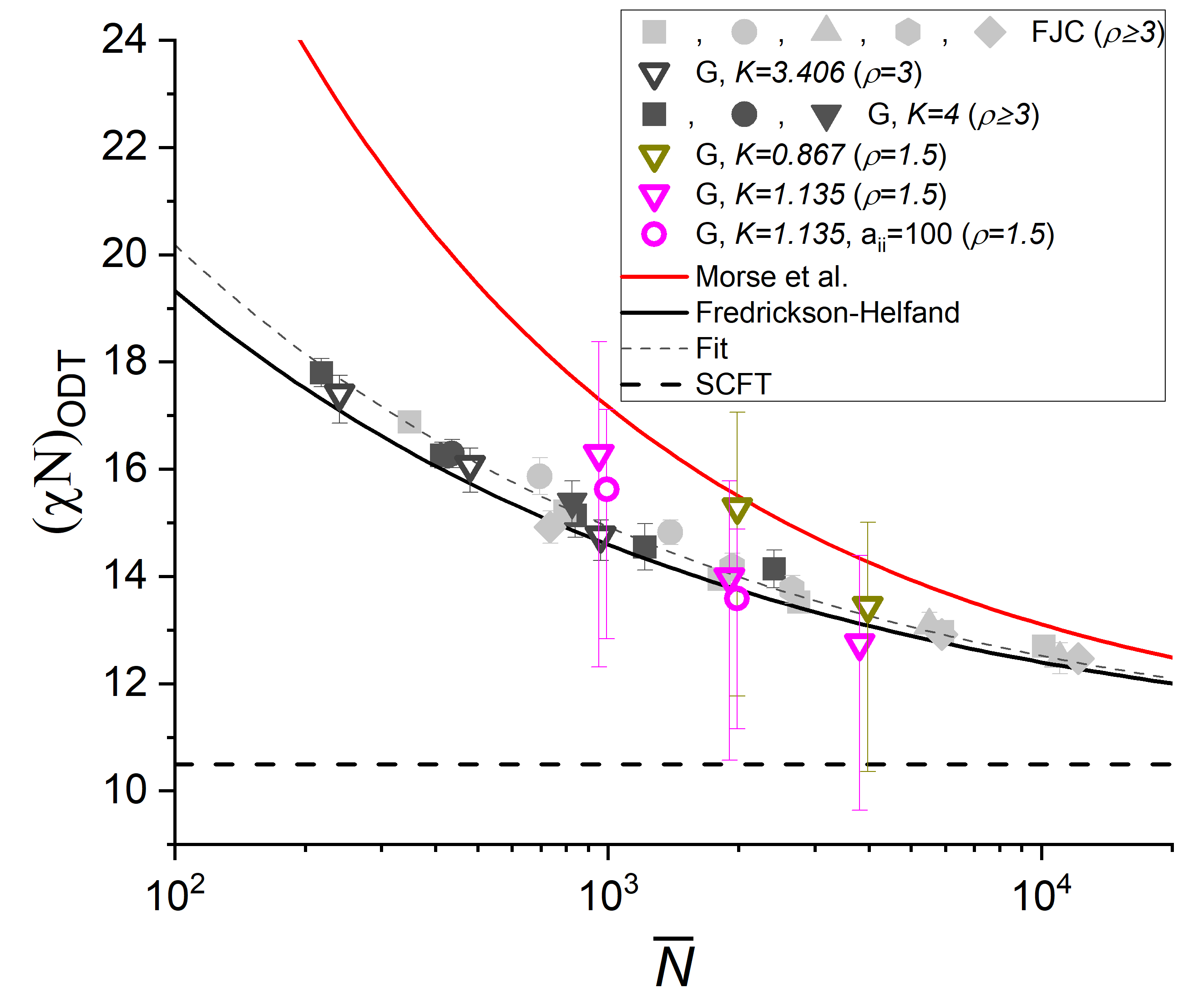}
  \caption{$(\chi N)_\text{ODT}(\bar{N})$ dependencies. The models having near-symmetric $z(N)$-distributions are shown in gray, colored symbols show the ODT points for the models with skewed $z(N)$-distributions. Table \ref{table2} contains correspondence between the symbol shape and the set of model parameters. For Gaussian (G) models with $K\leq 3.406$, we used the median value of $z(N)$-distributions to calculate $z$ via the procedure outlined in Supporting Information, section 2. The mean and median of the $z(N)$-distributions in all other models virtually coincided, and therefore we did not recalculate $z$ using the median value for those models. For G models with $K\leq 3.406$, the upper and lower limits of the error bars denote the $(\chi N)_\text{ODT}$ values obtained using the mean and the mode of the $z(N)$-distributions, respectively. Red, black, and thick dashed curves represent the predictions of Eqs. 2 and 1 in the main text and SCFT, respectively.}
  \label{fig:s_median}
\end{figure}

The homogeneous melt was prepared by placing the necessary number of diblock copolymer chains in the box and simulating the system at $\alpha=0$ for $2\times 10^6$ DPD time steps. Fig. \ref{fig:s1}a confirms that the Gaussian statistics was reached in the melt as expected in an equilibrium system.

The initially ordered lamellar structure for determining $\chi_{dis}$ was obtained by spontaneous ordering of the homogeneous melt at the values of $\chi N$ in the vicinity of the ODT. In particular, the value of $\chi N$ at which this structure was obtained was set to (i) produce defectless lamellar structure and (ii) be $\chi N <\chi_{or} N+1$ to remain at the vicinity of ODT and produce the weakly phase separated structure (Fig. \ref{fig:s1}b).

\subsection{The Effect of the Fixed Box Side Length on $(\chi N)_\text{ODT}$}

In this section, we will discuss the influence of the fixed box size on the position of the obtained ODT point. The finite fixed size of the simulation box can affect the value of $\chi_\text{ODT}$ when the box size is small enough, and the incommensurability of the side length and the lamellar period starts to affect ordering \cite{medapuram2015universal}. To confirm that the finite-size effects did not influence our results strongly, we carried out the following two procedures. 

First, we measured the $\chi_\text{ODT}$ value for a system having twice as many particles as the original system, similarly to ref. \cite{gavrilov2023effect}. To perform this check, we chose the system that has small number of lamellar periods and a box size incommensurate with the lamellar period. As a result, we chose the FJC model with $\rho=5$ and $r_0=1.26$ (see Table \ref{table2}). For the box side length $L=30$, the box contained 4 lamellar periods at ODT (Fig. \ref{fig:s1}b). We increased the box side length to $L=37.8$ to double the number of chains in the box and performed calculations of $\chi_{or}$ and $\chi_{dis}$. First, we observed that in the larger box, the number of lamellar periods increased to 5 (Fig. \ref{fig:s1}b). Second, we established that $\chi_{dis}$ coincided for both systems (the values of $\chi$ were changed with a step size $0.1/N=0.005$). Third, we observed an increase of $\chi_{or}$ by 0.01 upon the increase of $L$. Therefore, the $(\chi N)_\text{ODT}$ value increased by 0.1 when the system size doubled, which is well within the uncertainty associated with the difference between $\chi_{dis}$ and $\chi_{or}$ (Fig. 2a in the main text). This effect is expected to be even smaller for the systems with larger $\#_{Lam}$. Therefore, we concluded that performing the simulations in NVT ensemble did not affect the values of $(\chi N)_\text{ODT}$ in the systems with $\#_{Lam}\geq 4$ notably.

Second, to confirm the negligible influence of the fixed simulation box side length on $\chi_\text{ODT}$ in all studied models, we performed NPT simulations of the following two models: (i) FJC model with $\rho=3$, $N=20$, $r_0=1$ ($\#_{Lam}=5$) and (ii) Gaussian chain model with $K=1.135$, $\rho=1.5$, $N=16$ (the system with the smallest $\#_{Lam}$, $\#_{Lam}=3$). We used Berendsen barostat to maintain target system density at a given $\chi N$. All three dimensions of the rectangular simulation box were fluctuating independently. The value of $(\chi N)_\text{ODT}$ obtained in the NPT simulations coincided with the value observed in NVT simulations, which confirmed that the fixed simulation box side length affected the ODT point location negligibly.

\subsection{Additional Data}

\subsection{The Mapping of FJC Models Onto Atomistic Models of Block Copolymers}

\begin{figure}
\centering
  \includegraphics[width=\linewidth]{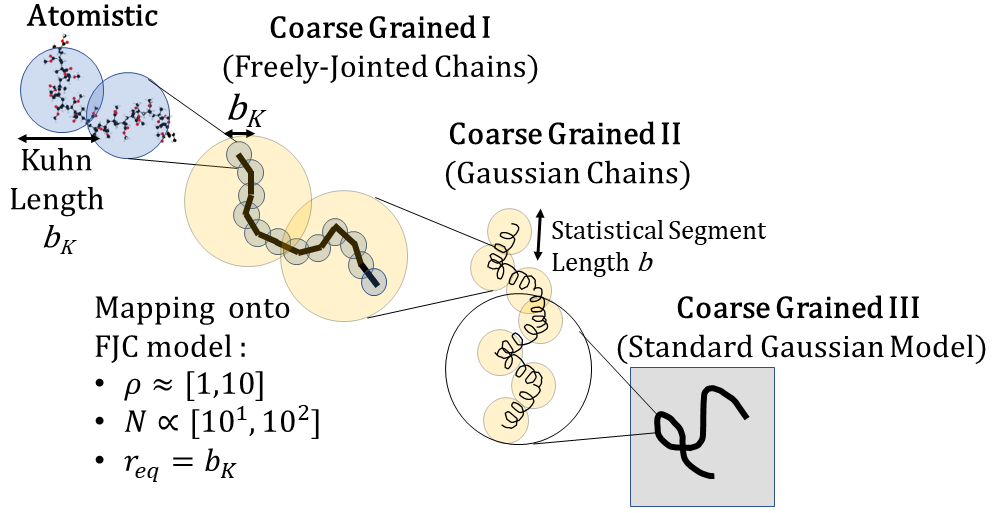}
  \caption{Sketch of the coarse-graining procedure. The direct mapping of the atomistic model onto FJC model is possible by choosing the equilibrium bond length equal to $b_K$ and by setting the values of $\rho$ and $N$ well accessible in simulations.}
  \label{fig:11}
\end{figure}

We discovered that one of the
groups of polymer models behaving universally at ODT in the $\chi\propto \alpha$ approximation are the freely-jointed chain (FJC) models with $\rho\geq 3$ (Fig. 2a in the main text). These models differ from the Gaussian chain models in one subtle detail: the bond length between beads is fixed and equal to a nonzero target bond length $r_0$ (or, in practice, bond length fluctuations are greatly suppressed \cite{huang2019dissipative}). Interestingly, there exists a direct mapping between FJC models and the atomistic models of the real diblock copolymers. Indeed, an equilibrium bond length $r_{eq}$ in an FJC naturally equals to the Kuhn length of a real polymer chain (Fig. \ref{fig:11}). Typical diblock copolymers consist of $\approx 10-100$ Kuhn segments \cite{sinturel2015high}, which is similar to the number of beads in a chain $N$ accessible to coarse-grained simulations \cite{glaser2014universality}. Finally, to match the experimental polymer density, the model density $\rho$ should be around $\rho\approx 1-10$, which is also well achievable in simulations (Table \ref{table2}). Further coarse-graining (Fig. \ref{fig:11}), if done consistently, requires smaller $N$, yet the density $\rho$ becomes computationally inaccessible if one is trying to map the model to the original underlying system. Therefore, the coarse-grained FJC models represent an important class of models having a direct correspondence to atomistic models of block copolymers.

\bibliography{sample}
\end{document}